\documentclass{IEEEtran}
\usepackage{mathtools}
\usepackage{cite}
\usepackage{amsmath,amssymb,amsfonts}
\usepackage{amsthm} 
\usepackage{graphicx}
\usepackage{textcomp}
\usepackage{subcaption}
\usepackage[algo2e,ruled, vlined, linesnumbered]{algorithm2e}
\usepackage{comment}
\usepackage{longtable}
\usepackage{xcolor}
\usepackage{algorithm}
\usepackage{algorithmic}
\usepackage{hyperref}
\usepackage{graphicx}
\usepackage{array}
\usepackage{booktabs}
\usepackage{multirow}
\hypersetup{colorlinks,linkcolor={blue},citecolor={blue},urlcolor={red}}
\usepackage[capitalise,nameinlink]{cleveref}

\usepackage[algo2e,ruled, vlined, linesnumbered]{algorithm2e}

\usepackage{amsmath}
\usepackage{graphicx}
\usepackage{longtable}
\usepackage{array}
\usepackage{booktabs}
\usepackage{multirow}
\usepackage{caption}
\usepackage{xcolor}
\usepackage{pifont}
\usepackage{colortbl}
\usepackage{tikz}
\usepackage{cuted}
\usetikzlibrary{shapes.geometric, arrows, shadows, positioning}

\begin{document}
	\title{\LARGE Energy Efficiency Maximization for CR-NOMA
		based Smart Grid Communication Network}
	\author{\noindent Mubashar Sarfraz,  Sheraz Alam , Sajjad A. Ghauri, Asad Mahmood 
		\thanks{ \noindent Mubashar Sarfraz \& Sheraz Alam are with the Department of Electrical Engineering, NUML, Islamabad, Pakistan email: mubashar.sarfraz@numl.edu.pk; salam@numl.edu.pk,\\
        \noindent Sajjad A. Ghauri is with the School of Engineering \& Applied Sciences, ISRA University, Islamabad, Pakistan email: dr.saghauri@gmail.com,\\
        \noindent Asad Mahmood is with the Interdisciplinary Centre for Security,
Reliability and Trust (SnT), University of Luxembourg, 4365 Luxembourg
City, Luxembourg. Email: asad.mahmood@uni.lu.
		}	
	}
	\maketitle
\begin{abstract}
Managing massive data flows effectively and resolving spectrum shortages are two challenges that Smart Grid Communication Networks (SGCN) must overcome. To address these problems, we provide a combined optimization approach that makes use of Cognitive Radio (CR) and Non-Orthogonal Multiple Access (NOMA) technologies. Our work focuses on using user pairing (UP) and power allocation (PA) techniques to maximize energy efficiency (EE) in SGCN, particularly within Neighbourhood Area Networks (NANs). We develop a joint optimization problem that takes into account the real-world limitations of a CR-NOMA setting. This problem is NP-hard, nonlinear, and nonconvex by nature. 
To address the computational complexity of the problem, we use the Block Coordinate Descent (BCD) method, which breaks the problem into UP and PA subproblems. Initially, we proposed the Zebra-Optimization User Pairing (ZOUP) algorithm to tackle the UP problem, which outperforms both Orthogonal Multiple Access (OMA) and non-optimized NOMA (UPWO) by 78.8\% and 13.6\%, respectively, at a SNR of 15 dB. Based on the ZOUP pairs, we subsequently proposed the PA approach, i.e., ZOUPPA, which significantly outperforms UPWO and ZOUP by 53.2\% and 25.4\%, respectively, at an SNR of 15 dB. A detailed analysis of key parameters, including varying SNRs, power allocation constants, path loss exponents, user density, channel availability, and coverage radius, underscores the superiority of our approach. By facilitating the effective use of communication resources in SGCN, our research opens the door to more intelligent and energy-efficient grid systems. Our work tackles important issues in SGCN and lays the groundwork for future developments in smart grid communication technologies by combining modern optimization approaches with CR-NOMA.
\end{abstract}
\begin{IEEEkeywords}
Smart Grid Communications, Non-orthogonal multiple access, Cognitive radio, Energy efficiency, Zebra optimization algorithm.
\end{IEEEkeywords}
\vspace{-1mm}
\section{Introduction}
Since the installation of the conventional grid, the world’s energy consumption has increased progressively \cite{deng2014residential}. The increasing demand for electricity has led to blackouts \cite{yan2013multi}, high prices, poor power quality, and environmental damage \cite{refaat2021smart, vaiman2011risk}. However, developed nations are taking necessary measures to introduce renewable energy sources and improve their electric grids to ensure reliable, efficient, and sustainable power supply. To meet the rapidly growing demand, governments and energy providers have improved energy and demand side management (DSM) initiatives \cite{bahrami2017online}. For a couple of decades, policies that allow increased distributed generation have been in place. In addition to distributed generation initiatives, the utilization of renewable energy sources has also grown steadily over time \cite{kabalci2018wireless,fakhar2023smart,rehmani2018integrating}. 
\par 
Traditional grids have evolved into SG through improvements in physical design and resource management. Efficient management of load and source \cite{kanakadhurga2022demand}, precise tracking of production and consumption rates \cite{lin2020novel}, and strategic implementation of control mechanisms \cite{mbungu2023control} are crucial factors in optimizing SG performance. The Internet of Energy combines wireless sensor networks, smart meters, actuators, and other components of the power grid with information and communication technologies \cite{fouda2011lightweight}. This has led to the need for communication-based networks as a prerequisite for SG. This technology predicts future actions to increase EE and reduce costs using bidirectional communication within the SG. 
\par
\begin{figure}[tbph!]
\centering
\includegraphics[width=0.9\linewidth]{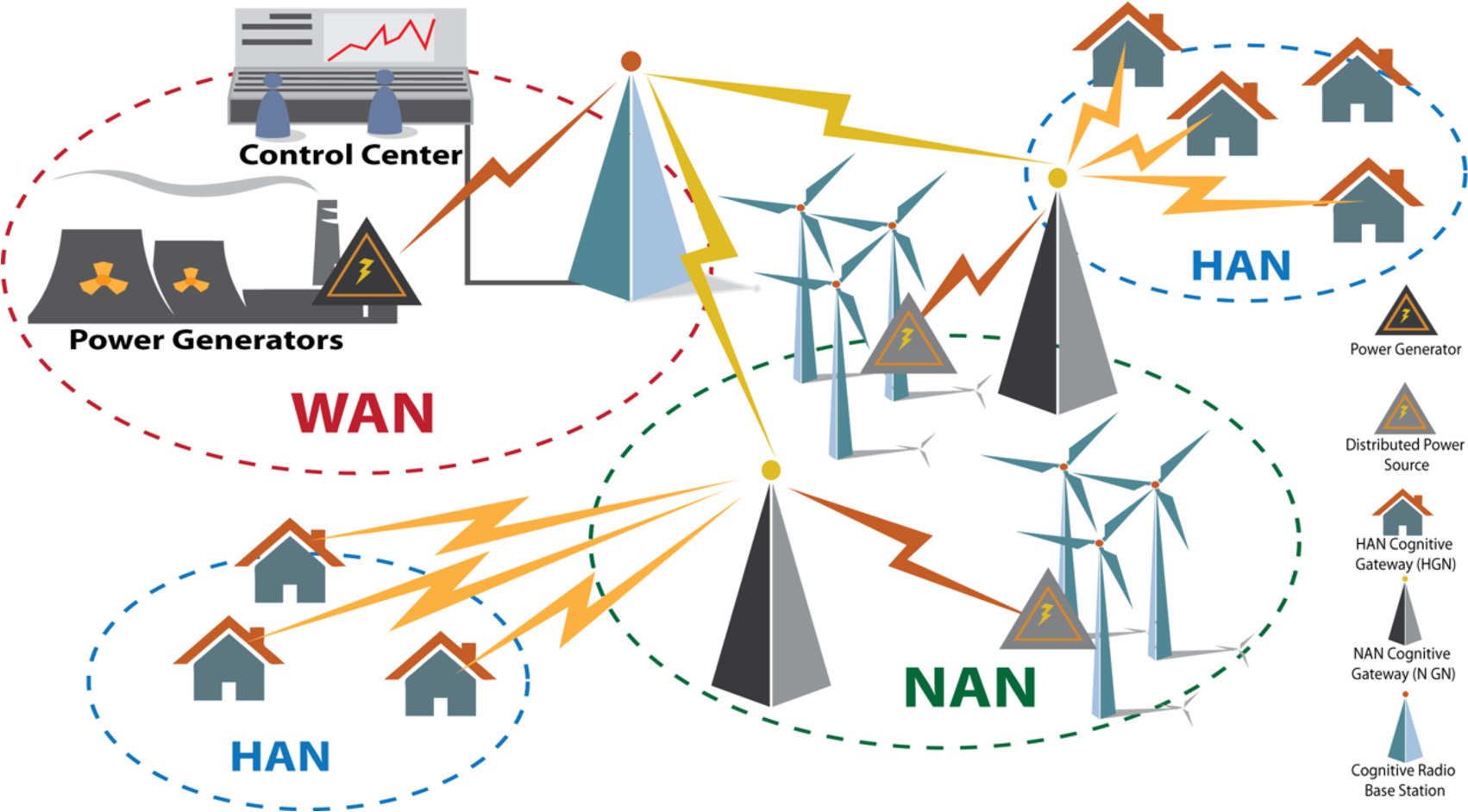}
\caption{Layered SGCN architecture}
\label{fig1} \vspace{-2mm}
\end{figure}

A substantial amount of data is generated using smart sensor networks for real-time monitoring of electricity generation, distribution, and consumption \cite{wang2017wireless,daki2017big,tu2017big,syed2020smart,ma2013smart,faheem2019software}. The data generated by these devices must be communicated to the control center, where necessary actions will be taken. Figure \ref{fig1} illustrates the layered architecture of the SG communication network (SGCN); home area network (HAN), neighborhood area network (NAN), and wide area network (WAN)\cite{zhang2013trust}. 
The HAN comprises smart devices and sensors installed in the home that are responsible for controlling and collecting data from the home network. In every home, smart meters (SMs) are installed to monitor DSM and metering data. This data is then transmitted to the NAN, where the data collector (DC) is stationed. The DC collects data from multiple SMs and sensors that monitor electricity distribution and renewable energy generation. The data from the multiple DCs is transmitted to the WAN and ultimately arrives at the control center, where appropriate actions are promptly implemented. Every application utilizing these networks must be allocated a designated amount of bandwidth, a precise level of latency, a traffic model, and a priority for transmitting data \cite{ghassemi2010cognitive}-\cite{ chaudhary2018sdn}.
\par 
The SGCN generates a significant amount of diverse data, necessitating a broader spectrum and higher power resources. The varied data rate needs and coverage ranges for HAN, NAN, and WAN architectures \cite{kuzlu2014communication} underscore the potential of CR technology to fulfill the spectrum requirements of different SGCN networks. Designing and implementing an SGCN is challenging because it requires merging various communication network segments that connect heterogeneous devices dispersed across enormous distances with variable quality of service (QoS) needs.  CR-based SGCNs are becoming important due to the need to fulfill the various requirements of varied architectures \cite{khan2015cognitive}. CR empowers dynamic spectrum access networks to effectively utilize spectrum opportunistically without interfering with primary users.

Similarly, a higher amount of power resources is necessary to distribute power to the large number of sensors in SGCN, creating a demand for energy efficiency (EE) in the smart grid. NOMA has emerged as a promising spectrum access technique that maximizes the system's EE \cite{fs1}.
An effective strategy to enhance the EE and environmental sustainability of the SGCN is through the implementation of CR-NOMA for power allocation and channel resource management.
\par
\vspace{-3mm}
\subsection{Related work}
This section discusses the existing literature utilizing NOMA to improve EE and the integration of CR with the SGCNs to meet spectrum demands. 
\par
For SGCN, a CR is proposed to place a power price on the efficiently received traffic data in a meter management system to collect data \cite{yang2016optimisation}. 
A wideband hybrid access strategy is proposed and analyzed to share the spectrum between SG nodes and CR networks and minimize system power costs. The authors optimized the sensing and transmission time while considering interference to primary users and loss of spectrum opportunity for secondary users. The authors in \cite{huang2013priority} analyzed the multimedia traffic of the SG systems using the CR communication infrastructure and proposed a priority-based approach to scheduling traffic. This approach is based on the different types of traffic that SG may generate, such as control commands, multimedia sensing data, and readings of the meters. Similarly, the authors of \cite{yu2015qos} presented differential QoS scheduling in CR-based SGCNs. The scheduler manages and allocates the available spectrum resources while arranging the SG users' data transmissions.
\par

The work presented in \cite{rabie2018two} introduced a dual-hop cooperative power line communication system based on NOMA, aiming to enhance system capacity by employing NOMA at the source and relay nodes. Similarly, the authors in \cite{rabie2017comparison} presented a NOMA-based decode and forward relaying for power line communication to maximize throughput and enhance system fairness.
\par
In \cite{fs1}, authors optimized EE for NOMA-based unmanned aerial vehicles (UAVs) by determining the optimal altitude of the UAV and power allocation through the nested Dicklebach structure. A fixed transmission power is taken as the starting point and an optimal altitude of $H_{o}$ is determined. The optimal transmission power is calculated using the $H_{o}$ to meet the minimum QoS requirements. Based on the results, it was found that maintaining a minimum altitude results in higher EE while still meeting the minimum QoS requirements, which can save 18\% of power. During hovering, the signal transmission process can conserve 49\% of the power.
The authors of \cite{fs1} extended the EE optimization problem and presented user pairing problems along with the altitude and power allocation factor problems in \cite{fs2}. The near-optimal user pairing is determined through a cat swarm optimization algorithm. The cat swarm optimization algorithm is used to determine the near-optimal user pairing. This work takes into account the power used for signal transmission, but does not consider the power used for hovering UAVs. Finally, reducing the altitude of the UAV can save 20\% of transmission power.
\par

The work in \cite{zamani2020optimizing} proposed EE for the uplink and downlink scenarios using the Dinkelbach algorithm in combination with the ellipsoid method and epigraph followed by successive convex approximation, which transforms the nonconvex problem into a sum of parameterized problems. This approach transforms the original non-convex problem into a weighted-sum EE problem. In \cite{al2021ee}, the authors optimized EE for downlink NOMA in cloud radio networks using stochastic geometry. The subgradient and false position methods are proposed for the optimal and suboptimal power allocation schemes, respectively.

\par
The authors of \cite{wu2021energy} proposed that the uplink CR-NOMA optimizes the system's EE using the Dinkelbach and Lagrange algorithm. The authors assessed the acceptable range of power usage ratios by considering each user's minimum QoS requirements.
The actor-critic reinforcement learning technique proposed in \cite{liang2020energy} enables a cluster of CR users to access the same spectrum simultaneously to improve the EE of CRNs. The weighted data rate serves as the reward function for this process, while the power allocation action strategy undergoes continuous evaluation and adjustment.
\par
In \cite{zhao2019energy}, the EE for the CR-NOMA was maximized subject to harvested energy and the QoS of the users. Optimizing a multi-objective optimization problem can be achieved by breaking it down into three subproblems: transmission optimization, power allocation, and power splitting ratio optimization. The Lagrange dual algorithm based on the first-order Taylor series expansion function is proposed to optimize the transmit power. Additionally, a multi-objective iterative algorithm has been introduced to attain a combined optimal solution.

\par
A power allocation scheme was proposed for CR-NOMA \cite{liang2020energy} to improve the energy self-sustainability of the CU so that many CU may be served simultaneously. An EE based on NOMA was used in the mobile edge cloud for optimal task scheduling among the mobile edge cloud servers proposed in \cite{dong2020noma}. The work in \cite{hua2018power} presented that the sensor node's UAV scheduling, PA scheme, and flight trajectory are designed to maximize the EE of the UAV by satisfying the user's QoS requirements. 
\par
In \cite{masood2021energy}, the smart meters utilizing IoT technology were thoroughly examined, and a pragmatic solution was put forth for efficient power distribution in an OFDM-DAS setup that employs distributed antennas. 
EE optimization in a smart grid system is achieved using OFDM-DAS technology, while IoT-powered SWIPT technology allows smart meters to collect energy and transmit data efficiently. The proposed solution optimizes power allocation for an SG and increases the functionality of smart meters with SWIPT capability. The objective function considers three options: limiting EH, adjusting the PS ratio, and managing the DAS transmit power. The nonconvex EE maximization problem is solved using nonlinear fractional programming and the Lagrangian approach.
\par
The authors of \cite{asuhaimi2019channel} combined cellular networks and SGCN to optimize EE and minimize delays in the NAN environment. The authors proposed a scheme for accessing channels and controlling power distribution and developed a learning-based approach for phasor measurement units (PMUs) to transmit data while considering interference constraints. The PMUs are trained through reinforcement learning to identify the optimal strategy for achieving maximum successful transmissions, regardless of their prior knowledge of the system's dynamics. 
\par
The authors of \cite{lin2021secrecy} explored secrecy EE in hybrid beamforming for a satellite-terrestrial integrated network, where a multibeam satellite shares the millimeter wave spectrum with a cellular system. They addressed the challenges posed by imperfect angles of departure in wiretap channels and designed a hybrid beamformer at the base station, along with digital beamformers for the satellite, to maximize secrecy EE while ensuring that the SINR constraints were met for both terrestrial and cellular users. The problem was non-convex, so the authors utilized the Charnes-Cooper approach and an iterative search method to reformulate the problem for determining the beamforming weights.
\par
Multicast communication in a satellite and aerial integrated network using rate splitting multiple access was examined in \cite{lin2021supporting} to improve spectral efficiency and reduce hardware complexity, along with a new beamforming scheme. The network included both satellite and UAV components operating in the same frequency band, managed by a central network management center, focusing on efficient content delivery and interference suppression for numerous IoT devices. The formulated problem was non-convex and s solution to the optimization problem through sequential convex approximation and iterative methods is presented, aimed at maximizing data transmission rates within SINR and power constraints.
\par
The authors in \cite{lin2022refracting} proposed a method to enhance communication in hybrid satellite-terrestrial networks with blocked users by using a refracting RIS. A joint beamforming design was presented where a BS acts as a half-duplex decode-and-forward relay, aiming to minimize total transmit power while meeting user rate requirements. The optimization problem was complicated by the interdependence of beamforming weight vectors and RIS phase shifters. To solve this, the authors presented a singular value decomposition scheme and uplink-downlink duality for beamforming, along with Taylor expansion and penalty function methods for phase shifting. 
\par
The work in \cite{an2024exploiting} developed a multi-layer refracting RIS receiver architecture for high-altitude platform-assisted simultaneous wireless information and power transfer networks. This architecture enhances energy transfer while reducing fading effects in long-distance communications. The authors formulated a problem aimed at maximizing the worst-case sum rate, taking into account channel imperfections and energy harvesting constraints. To address this problem, they proposed a scalable and robust optimization framework. This framework included a discretization method, the LogSumExp dual algorithm, and a modified cyclic coordinate descent approach.
\par
Table \ref{tab2}, provides a concise summary of the relevant literature. Researchers have focused on CR-based SGCN, improving EE in NOMA by looking at factors like user pairing, power allocation, and channel allocation. There is still a need to explore the use of CR-NOMA to enhance the EE of the SGCN.

\begin{table*}[h!]
\caption{Literature Review of Existing Techniques}
\label{tab2}
\small
\centering
\begin{tabular}{llllllllll} \hline
\textbf{Ref.} &\textbf{OF} & \textbf{UP} & \textbf{PA} & \textbf{CA} & \hspace{0.25mm} \textbf{FT} & \textbf{CR} & \textbf{SG} & \textbf{Application Domain}  \\ \hline
\cite{fs1}  & EE  &  \hspace{0.3mm}    -      &   \checkmark & \hspace{0.3mm} -    &\hspace{1mm}   \checkmark       &  \hspace{1mm}    -       &\hspace{1mm}   - &   \hspace{1mm}      Aerial NOMA    \\ \hline
\cite{yang2016optimisation}  & ST, TT  &  \hspace{0.3mm}    -      &   \hspace{0.3mm} - & \hspace{0.3mm} -    &\hspace{1mm}   \hspace{0.3mm} -       & \hspace{0.3mm}  \checkmark       &\hspace{0.3mm}  \checkmark &   \hspace{6mm}      MDMS    \\ \hline
\cite{huang2013priority}  & SUO  &  \hspace{0.3mm}    -      &   \hspace{0.3mm} - & \hspace{0.3mm} \checkmark    &\hspace{1mm}   \hspace{0.3mm} -       & \hspace{0.3mm}  \checkmark       &\hspace{0.3mm}  \checkmark &   \hspace{6mm}      SGCN    \\ \hline
\cite{yu2015qos}  & Delay  &  \hspace{0.3mm}    -      &   \hspace{0.3mm} - & \hspace{0.3mm} \checkmark    &\hspace{1mm}   \hspace{0.3mm} -       & \hspace{0.3mm}  \checkmark       &\hspace{0.3mm}  \checkmark &   \hspace{6mm}      SGUs    \\ \hline

\cite{fs2}   & EE  & \hspace{0.1mm}   \checkmark     &    \checkmark  & \hspace{0.3mm} -    &\hspace{1mm}   \checkmark       &  \hspace{1mm}    -       &\hspace{1mm}   - &   \hspace{1mm}      Aerial NOMA    \\ \hline
\cite{zamani2020optimizing}  & EE  &\hspace{0.3mm}   -    &    \checkmark     &\hspace{0.3mm}   -  & \hspace{0.3mm} -
 &\hspace{1mm}   -        &\hspace{1mm}   -     & \hspace{8mm} ND            \\ \hline
\cite{al2021ee}    & EE, Sum-rate      &\hspace{0.3mm}  -   &  \checkmark   & \hspace{0.3mm} -  &\hspace{1mm}    -    &\hspace{1mm}  -   &\hspace{1mm}   -    &  \hspace{1mm} Cloud RAN            \\ \hline
\cite{wu2021energy}    & EE      &\hspace{0.3mm}   -   &  \checkmark & \hspace{0.3mm} -     &\hspace{1mm}   -    &\hspace{1mm}   \checkmark    &\hspace{1mm}   -    & \hspace{8mm} IoT           \\ \hline
\cite{liang2020energy}    & EE, SE      &\hspace{0.3mm}   -   & \checkmark & \hspace{0.3mm} -     &\hspace{1mm}   -    &\hspace{1mm}   \checkmark    &\hspace{1mm}   -    & \hspace{8mm} ND           \\ \hline
\cite{zhao2019energy}    & EE      &\hspace{0.3mm}   -   & \checkmark & \hspace{0.3mm} -   &\hspace{1mm}   -    &\hspace{1mm}   \checkmark    &\hspace{1mm}   -    & \hspace{8mm} ND           \\ \hline
\cite{dong2020noma}    & EE      &\hspace{0.3mm}   -   & \checkmark  & \hspace{0.3mm} -  &\hspace{1mm}   -    &\hspace{1mm}   -   &\hspace{1mm}   -    & \hspace{7mm} MEC           \\ \hline
\cite{hua2018power}    & TP      &\hspace{0.3mm}   -   &  \checkmark     & \hspace{0.3mm} - &\hspace{1mm}    \checkmark    &\hspace{1mm}   -   &\hspace{1mm}   -    &  UAV-based WSN           \\ \hline
\cite{masood2021energy}    & EE      &\hspace{0.3mm}   -   &   \checkmark      & \hspace{0.3mm} - &\hspace{1mm}    -    &\hspace{1mm}   -   &\hspace{1mm}   \checkmark    &  IoT enabled SMs           \\ \hline
\cite{asuhaimi2019channel}    & EE, Delay      &\hspace{0.3mm}   -   &  \checkmark   & \hspace{0.3mm} \checkmark  &\hspace{1mm}    -    &\hspace{1mm}   -   &\hspace{1mm}   \checkmark    &  \hspace{1mm} NAN in SGCN            \\ \hline
\cite{alam2018clustering}  & Fairness, UR  &  \hspace{0.3mm}    -      &   \hspace{0.3mm} - & \hspace{0.3mm} \checkmark     &\hspace{1mm}   \hspace{0.3mm} -       & \hspace{0.3mm}  \checkmark       &\hspace{0.3mm}  \checkmark &   \hspace{1mm} NAN in SGCN   \\ \hline
\cite{alam2017dynamic}  & Fairness, UR  &  \hspace{0.3mm}    -      &   \hspace{0.3mm} - & \hspace{0.3mm} \checkmark     &\hspace{1mm}   \hspace{0.3mm} -       & \hspace{0.3mm}  \checkmark       &\hspace{0.3mm}  \checkmark &   \hspace{1mm} NAN in SGCN   \\ \hline
\cite{alam2019joint}  & EE, Fairness, UR  &  \hspace{0.3mm}    -      &   \hspace{0.3mm} \checkmark & \hspace{0.3mm} \checkmark     &\hspace{1mm}   \hspace{0.3mm} -       & \hspace{0.3mm}  \checkmark       &\hspace{0.3mm}  \checkmark &   \hspace{1mm} NAN in SGCN   \\ \hline

\cite{jayachandran2022power}    & DR, Latency      &\hspace{0.3mm}   -   &  \checkmark   & \hspace{0.3mm} -  &\hspace{1mm}    -    &\hspace{1mm}   -   &\hspace{1mm}   \checkmark    &  \hspace{1mm} SDs in SGCN            \\ \hline
\cite{hussain2020performance}    & SE      &\hspace{0.3mm}   -   &  \checkmark   & \hspace{0.3mm} -  &\hspace{1mm}    -    &\hspace{1mm}   -   &\hspace{1mm}   \checkmark    &  \hspace{1mm} SMs in SGCN            \\ \hline
\cite{faheem2019multi}    & Capacity, SpU      &\hspace{0.3mm}   -   &  -   & \hspace{0.3mm} \checkmark  &\hspace{1mm}   -    &\hspace{1mm}   \checkmark   &\hspace{1mm}   \checkmark    &  \hspace{1mm} CRSNs-based SG            \\ \hline
\textbf{OW}    & EE    &\hspace{0.3mm}    \checkmark      &  \checkmark & \hspace{0.3mm} \checkmark     &\hspace{1mm}   -   &\hspace{1mm}    \checkmark  &\hspace{1mm}   \checkmark    & \hspace{1mm} NAN in SGCN              \\ \hline 
\end{tabular}
 	 \\ \vspace{1.5mm} \noindent{\footnotesize{UP; User Pairing in NOMA, PA; Power Allocation, FT; Flight Trajectory, CA; Channel Access, TP; Transmit Power, ST; Sensing Time, TT; Transmission Time, SM; Smart Meters, MEC; Mobile Edge Cloud, ND; Not Defined, RAN; Radio Access Network, MDMS; Metre Data Management System, SUO; System Utility Optimization, DR; Data Rate, SpU; Spectrum Utlization, CRSNs; Cognitive Radio Sensor Netwokrs UR; User Reward SGUs; SG Users, SE; Spectral Efficiency, SDs; Sensor Devices, OF; Objective Function, \textbf{OW}; Our Work}}
\end{table*}
\vspace{-2mm}
\subsection{Motivation and Contributions of the paper}
Based on the existing literature, NOMA has been proposed as a means to enhance EE, spectral efficiency, and throughput across a range of applications, as given in \cite{fs1,fs2,zamani2020optimizing,al2021ee,wu2021energy,liang2020energy,zhao2019energy,dong2020noma,hua2018power}. The NOMA-based approaches optimize power allocation, user pairing, altitude optimization, or a combination of these approaches. Similarly, the authors introduced cognitive radio and channel allocation in SGCN to address spectrum scarcity, enhance user fairness, and maximize the throughput of the SG \cite{alam2018clustering,alam2017dynamic,alam2019joint,yang2016optimisation,huang2013priority,masood2021energy,asuhaimi2019channel,jayachandran2022power,hussain2020performance,faheem2019multi}. The authors utilize channel allocation or cognitive radio in SGCN to effectively optimize the desired utility function. However, the literature review highlights a research gap in which work has yet to be done to utilize NOMA in SGCN for the simultaneous optimization of user pairing, power allocation, and channel allocation using CR. 

\par 
\par 
To the best of the author's knowledge, research has not yet been conducted to integrate CR-NOMA into an SG communication scenario in order to meet communication requirements and improve the EE of the SG network. NAN-DC collects SM data and transmits it to the control center for processing. The transmission of the data from SM to NAN-DC requires a spectrum resource. 
Based on the communication model of \cite{alam2018clustering,alam2017dynamic,alam2019joint}, users can utilize CR to access various channels based on their geographical location. Considering the channel availability constraint, CR-NOMA can be effectively used to maximize the EE of the SGCN. Adherence to the NOMA principles allows us to pair users according to channel availability at the time of pairing and allocate power to achieve objectives.
\par
The key contributions of the paper are summarized as follows:
\begin{enumerate}
    \item We formulated a mathematical framework for joint UP and PA problems that combines CR and NOMA for the NAN scenario of SGCN. This considers the practical limitations associated with the simultaneous use of CR and NOMA in the SGCN. 
    Our approach maximizes resource allocation for spectrum management by leveraging CR-NOMA, in which CR addresses spectrum scarcity, and NOMA enhances resource efficiency by pairing users and dynamically assigning power to them.
    
    \item The joint optimization problem is a computationally complex, nonlinear, nonconvex, and NP-hard problem. 
    To address this, we have implemented a BCD method to decompose the joint optimization problem into two subproblems: UP and PA. In the first subproblem, we use CR and NOMA to allocate channels to user pairs. The second subproblem allocates power to the suboptimal user pairs to maximize EE.
    
 \item To solve the first sub-probelm, we proposed zebra optimization algorithm (ZOA) to assign channels to user pairs, considering the CR-NOMA constraints. In the subsequent phase, the ZOA is used to tackle the power allocation subproblem, which involves efficiently distributing power among the previously determined suboptimal 
 user pairs.

 \item The effectiveness of the joint optimization problem is assessed using extensive Monte Carlo simulations. In the first subproblem, we compared the performance of ZOUP with four benchmark schemes. The results indicated that ZOUP outperformed UPWO by 9.5\%, 16.32\% improvement for adjacent pairing, 20.21\% for random pairing, and 60.31\% for OMA at 40 dB of SNR. Furthermore, we evaluated the performance of ZOUPPA in comparison with various power allocation schemes, indicating that ZOUPPA achieved 18.34\% improvement over power allocation scheme of \cite{yang2016general} and 26.82\% over fixed power allocation at 40 dB SNR.
\end{enumerate}

\allowdisplaybreaks

The structure of this paper is as follows. Section \ref{SMPS} presents the system model and the joint optimization problem. Then, Section \ref{PS} describes the framework to find the solution. Section \ref{RD} discusses the numerical results, and finally, the paper is concluded in Section \ref{Con}.

\section{System Model and Proposed Solution} \label{SMPS}
\begin{figure}[tbph!]
\centering
\includegraphics[width=0.9\linewidth]{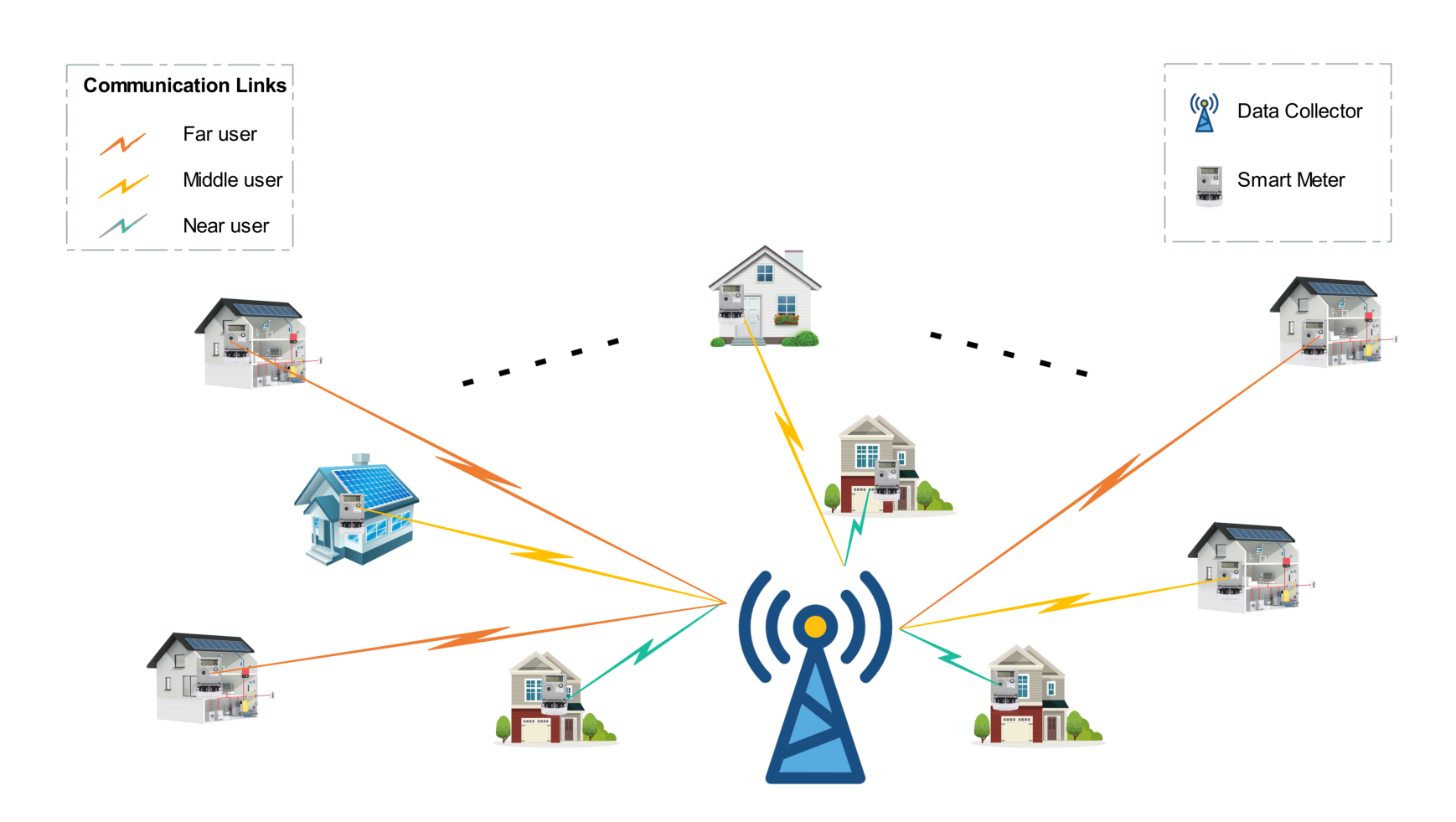}
\caption{NOMA-based communication model for NAN scenario.}
\label{fig3} 
\end{figure}

In this section, we discussed an SGCN where SMs gather data and send it to the control center through NAN-DC. Figure \ref{fig3} presents the proposed NOMA-based communication model. The classification of network users is based on the NOMA pairing criteria, which divide them into near, middle, and far categories. This work considers the assumption of \cite{alam2018clustering}, where the DC continually updates its database as the availability of the spectrum changes. In light of the open-loop regulatory paradigm, it can be assumed that the channel list will remain unchanged for at least 48 hours after the announcement.
The communication in CR-based SGs is designed to align with the IEEE 802.11af standard for utilizing TV white spaces. This approach is proposed within the context of the NAN architecture for SGCN as given in \cite{alam2019joint}. Given that metering data is generally less critical and can exhibit lower reliability, it is well-suited for transmission over a cognitive radio network \cite{alam2017cognitive}. In this framework, the SM acts as the secondary user, effectively leveraging the available spectrum to facilitate communication with the NAN-DC.
\vspace{-1mm}
\subsection{Mathematical Model} 
In this study, a CR-NOMA uplink system for SG communication is designed assuming that SM (users) are equipped with a single antenna. It is assumed that there are $N$ SMs deployed in a slow and varying suburban environment where the location and duration of $K$ channel availability do not change during single channel assignment.
NOMA users are divided into $M$ groups/clusters, each of which has access to a single resource block, and the users within a group transmit with different transmission power. 
Channel gains $g$ are calculated based on the assumption that the channels between DC and SM are Rayleigh fading channels and path loss for the suburban environment with path loss exponent $\chi$.
It is proposed that a maximum of two users in a cluster be paired together.
During the uplink, the NOMA pair transmits on the same channel, and the DC performs successive interference cancelation (SIC) to decode each user's signal. The transmitted signal of the $u^{th}$ user is given by Eq. \eqref{eq1} \vspace{-1mm}
\begin{equation}
\centering
x_{u}= \sqrt{p_{u}}s_{u},
\label{eq1} 
\end{equation}
where $p_{u}$ is the transmitted power and $s_{u}$ is the data signal of the $u^{th}$ user. The signal received by the DC from the $m^{th}$ cluster is given by \vspace{-2mm}
\begin{equation}
\centering
y_{m}= \sum_{u=1}^{U_{T}} g_{u}x_{u}+ \aleph_{m},
\label{eq2}
\end{equation}
where $UT$ is the total user number in a cluster, $g_u$ is the Rayleigh distributed fading channel coefficient, and $\aleph_m$ is the AWGN channel noise.
In OMA, the individual rate for the uplink transmission is defined as follows.
\begin{equation}
\centering
\gamma^{OMA}_{n}= \frac{1}{2} \log_{2} \left ( 1+ \frac {p_{n}}{\sigma^{2}} \vert g_{n} \vert^{2} \right),
\label{eq3}
\end{equation}
where $p_{n}$ is the maximum transmitted power, $\sigma^{2}$ is the noise power and $g_{n}$ is the channel gain of the $n^{th}$ user. 
\par
The DC can perform SIC on any of the users of the $m^{th}$ cluster. However, it is assumed that the DC will decode the signal of the $\nu^{th}$ (weak) user first, then the $\mu^{th}$ (strong) user. The uplink-NOMA transmission rate for the $\nu^{th}$ user and the $\mu^{th}$ user are represented as 
\begin{equation}
\centering
\gamma^{NOMA}_{\nu}=  \log_{2} \left ( 1+ \frac {\delta_{\nu} p}{\sigma^{2}} |g_{\nu}|^{2} \right) ,
\label{eq4}
\end{equation}
\begin{equation}
R^{\text{NOMA}}_{\mu} = \log_{2} \left( 1 + \frac{ \delta_{\mu} p |g_{\mu}|^{2} }{ \delta_{\nu} p |g_{\mu}|^{2} + \sigma^{2} } \right),
\label{eq5}
\end{equation}
where $\delta_{u}$ and $\delta_{v}$ are the power allocation factors for the $\mu^{th}$ and $\nu^{th}$ users, respectively. The objective is to maximize the EE of the CR-based SG communication system while ensuring that each user is subject to a minimum rate requirement. \vspace{-2mm}
\subsection{Problem Formulation}
The EE of the SGCN can be calculated by the following expression.
\begin{equation}
\begin{aligned}
\label{Equation6}
\eta_{EE}=  \sum\limits_{j = 1}^J \frac{ \left ( \gamma^{j}_{\mu} +\gamma^{j}_{\nu}\right)}{P_{j}} ,\\
\end{aligned}
\end{equation}
where $\gamma^{j}_{\mu}$ and $\gamma^{j}_{\nu}$ are the user rates $\mu$ and $\nu$ correspond to the $j^{th}$ group and $P_{j}$ is the total power assigned to a $j^{th}$ group. The optimization problem of EE for two-user NOMA can be formulated as follows: 
\begin{subequations}
\label{Equation116}
\begin{align}
\textbf{($\mathcal{P} 1$)}:  &\max_{{\textbf{\{$\boldsymbol{\Gamma}$, }\textbf{U, P}\}}} \eta_{EE} , \label{OF}\\
C_{1}:  \hspace{2mm} & \gamma^{NOMA}  \geq \gamma^{OMA} \quad \hspace{5mm} \forall  n, \\
C_{2}: \hspace{2mm} & p(\delta_{\mu} + \delta_{\nu}) \leq P_{j} \hspace{11mm} 0\leq P_{j} \leq 1 ,  \\
C_{3}: \hspace{2mm}  & \sum\limits_{j=1}^J P_{j} \leq P^{total} ,\\
C_{4}: \hspace{2mm} & \sum\limits_{l=1}^{U_{T}} u_{l} = 2 ,\\
C_{5}: \hspace{2mm} & \Gamma_{\mu, \nu}^m \in  [0,1], 
\quad \hspace{12mm} \forall m.
\end{align}
\end{subequations}
where \textbf{$\boldsymbol{\Gamma}$}, \textbf{U} and \textbf{P} are the decision variables, \textbf{$\boldsymbol{\Gamma}$ }is the channel availability matrix, \textbf{U} is a user pairing matrix and \textbf{P} is a power allocation vector. The minimum QoS of the NOMA user is provided through the $C_1$. The constraint $C_2$ defines the maximum usable power for each pair of users in a cluster $j^{th}$. $C_3$ ensures that the total power for all clusters should be under the total available power. The $C_4$ gives an upper limit to the maximum number of users in a cluster. The channel availability for each user is different, or all channels are not available for each user at a given time. Therefore, $C_5$ dictates that the user pairing must be performed so that when a channel is allocated, a user pair must have the same channel availability. So, $\Gamma^m$ means that the $m^{th}$ channel is available for the user pair $\mu$ and $\nu$ at a given time $t$.
\section{Framework for Solving Joint Optimization in MINLP} \label{PS}
The joint optimization problem mentioned in \eqref{Equation116} is non-linear due to the presence of the $\log(.)$ function in \eqref{eq4} and \eqref{eq5}. Furthermore, the fractional nature of the objective function in \eqref{OF} contributes to its non-convex nature. Moreover, the problem is mixed integer nonlinear programming (MINLP)  because of the discrete characteristics of user pairing and channel availability along with the continuous properties of the power allocation vector. In addition, coupling of the user pairing and power allocation with the channel availability matrix results in an NP-hard nature. Thus, to overcome the computational channeling of the joint optimization problem, we proposed a low complex solution by utilizing the block coordinate descent (BCD) approach, decoupling the joint optimization problem into a series of subproblems, and solving it iteratively to find the optimal (local) best solution \cite{mahmood2023joint}. Moreover, to narrow the gap between local optimal and global optimal is left for future work.  \vspace{-2mm} 
\subsection{Subproblem $\mathcal{P}$1-A}
The first subproblem for $\boldsymbol{\Gamma}$ and \textbf{U}, along with their associated constraints for fixed power allocation vector can be represented as:
\begin{subequations}
\label{Equation8}
\begin{align}
\textbf{($\mathcal{P}$1-A)}: & \max_{{\textbf{\{$\boldsymbol{\Gamma}$, U}\}}} \eta_{EE} \\ 
&\text{subject to:} ~ \text{C}_1 ~\text{to} ~\text{C}_5
\end{align}
\end{subequations}

The  \textbf{($\mathcal{P}$1-A)} is solved by considering two decision variables $\boldsymbol{\Gamma}$ and \textbf{U}, where user pairing is optimized through the zebra optimization algorithm (ZOA). Power allocation for \textbf{($\mathcal{P}$1-A)} is done by using the power allocation method of \cite{yang2016general}. The power allocation factor $\delta_u$ is derived from $C_1$ and by introducing two constant coefficients $\beta_1$ and $\beta_2$ it is expressed as: 
\begin{equation}
    \delta_u=\frac{\beta_1}{1+\sqrt{1+p{|g_u|}^2}} + \frac{\beta_2}{1+\sqrt{1+p{|g_v|}^2}},  \hspace{1mm}0\leq \beta_{r} \leq 1 \\
    \label{eq90a}
\end{equation}
and $\beta_1+\beta_2=1$. The power allocation for \textbf{($\mathcal{P}$1-A)} is carried out using eq. \eqref{eq90a} and power allocation for the weak user $v$ is $\delta_v=1-\delta_u$. Changing the constant value $\beta_2$  influences the data rate, and higher values achieve a higher data rate, resulting in the maximum EE of the system.

\subsubsection{ZOA based user pairing (ZOUP)}
The ZOA algorithm represents a new bioinspired metaheuristic algorithm based on the natural behavior of zebras. In ZOA, zebras are simulated as they forage for food and as they defend themselves against predators \cite{trojovska2022zebra}. Each zebra represents a candidate solution, that is, a possible channel allocation scheme for this problem.
\par 
The ZOA for the SGCN problem starts with initializing the parameters and the candidate solutions (zebras). The fitness of each zebra is calculated through \textbf{($\mathcal{P}$1-A)}, and the pioneer zebra (best candidate) is selected. The initial solution is updated in phase 1, i.e., foraging behavior, through the following equation:
\begin{algorithm2e}[b]
	\small
	\SetAlgoLined
	\textbf{Initialize }  $\boldsymbol{\Gamma}$, {\textbf{U}}, $SMs$, $M$, $\chi$, $SNR$, $R_{c}$, $Max\textunderscore Iterations$\\
\vspace{1mm}
{\textbf{get}} $\gamma^{OMA}_{n}$ from eq. \eqref{eq3}, Power allocation vector from eq. \eqref{eq90a}  \\
	\vspace{1mm}
	\While{iteration\textunderscore count $\textless$ Max\textunderscore Iterations }
{ 
\For{i=1:Clusters}{
\While{$\gamma^{NOMA}_{n}$ $\textless$ $\gamma^{OMA}_{n}$ }
{ // $\gamma^{NOMA}_{n}$ is a single user in a cluster (from eq. \eqref{eq4} or eq. \eqref{eq5})
\\
 \vspace{1mm}
 {\textbf{Apply Zebra Optimization to $\mathcal{P}$1-A}} eq. \eqref{Equation8} \\
 // subject to $C_{1}$ to $C_{5}$ of eq. \eqref{Equation116}
 \begin{enumerate}
 \vspace{1mm}
      \item Foraging Behavior
     \item Defence strategy against predators
 \end{enumerate}
	  {	
             }
             }
             \textbf{get} suboptimal pairing \\
              \eIf{$C_{1}$ to $C_{5}$ satisfied \& EE maximized}
  {\textbf{Exit} 
   \\
   }{
    \textbf{go} to step 7 \\
  }
    }
} 
\textbf{return } $\gamma^{NOMA}_{\nu}$, $\gamma^{NOMA}_{\mu}$
\caption{Zebra optimization-based user pairing (ZOUP)}
\label{algo1}
\end{algorithm2e}
\begin{equation}
\label{eq1010}
    \Gamma_i^{new}=\Gamma_i+\eta_{r} \times (\Gamma ^{best} -\omega \Gamma_i)
\end{equation}
where $\Gamma_i^{new}$ is the updated solution, $\Gamma_i$ is the initial solution, $\eta_{r}$ is the random number in the interval [0,1], $\Gamma ^{best}$ is the pioneer zebra, $\omega$ is the constant [1,2]. Therefore, a final solution will be selected after the greedy search, and the best candidate will be selected from the updated and initial solutions.
\par 
In phase 2, a defense strategy against predators is employed where two scenarios are considered: lion-attacked (case-1) zebras and other predators-attacked (case-2) zebras, and the solution is updated through the following equations:

\begin{algorithm2e}[t]
	\small
	\SetAlgoLined
	{\textbf{get}} Initial parameters $\&$ suboptimal pairing from \textbf{Algorithm \ref{algo1}}  \\
	\vspace{1mm}
	\While{iteration\textunderscore count $\textless$ Max\textunderscore Iterations }
{ 
\For{i=1:Clusters}{
\While{$\gamma^{NOMA}_{n}$ $\textless$ $\gamma^{OMA}_{n}$ }
{ // $\gamma^{NOMA}_{n}$ is a single user in a cluster (from eq. \eqref{eq4} or eq. \eqref{eq5})

 \vspace{1mm}
 {\textbf{Apply Zebra Optimization to $\mathcal{P}$1-B}} eq. \eqref{eq12} \\
 // subject to $C_{1}$ to $C_{3}$ of eq. \eqref{Equation116}
 \begin{enumerate}
 \vspace{1mm}
      \item Foraging Behavior
     \item Defence strategy against predators
 \end{enumerate}
	  {	
             }
             }
             \textbf{get} suboptimal power allocation vector \\
             \eIf{$C_{1}$ to $C_{3}$ satisfied \& EE maximized}
  {\textbf{Exit} 
   \\
   }{
    \textbf{go} to step 6 \\
  }
    }
} 
\textbf{return } $\gamma^{NOMA^{*}}_{\nu}$, $\gamma^{NOMA^{*}}_{\mu}$
 \caption{Zebra optimization-based joint user pairing and power allocation (ZOUPPA)}
\label{algo2}
\end{algorithm2e}

\begin{equation}
\label{eq11}
\Gamma_i^f=
\begin{dcases*}
\Gamma_i^{new} +\mathbb{R} (2\eta_{r}-1) (1- \frac{t_i}{T_{max}})\Gamma_i^{new} \hspace{2mm} \eta_{r} \leq 0.5 \\
\Gamma_i^{new} +\eta_{r} (\Check{\Check{A}}-\omega \Gamma_i^{new})   \hspace{20mm}   o.w
\end{dcases*}
\end{equation}

where $\Gamma_i^f$ is the final solution, $\mathbb{R}$ is the constant value and is taken as 0.1, $t_i$ is the current iteration, $T_{max}$ is the maximum number of iterations and $\Check{\Check{A}}$ is the status of the attacked zebra. The final solution is selected based on the greedy comparison between the updated and initial solutions. The ZOA for eq. \eqref{eq1010} to eq. \eqref{eq11} is repeated over $T_{max}$ iterations, and the best candidate among the updated is selected as the final solution. The proposed ZOUP to solve the maximization of EE in \textbf{($\mathcal{P}$1-A)} is shown in Algorithm \ref{algo1}.

\subsubsection{ZOA-based joint user pairing and power allocation (ZOUPPA)}

The joint user pairing and power allocation for the problem of \textbf{($\mathcal{P}$ 1)} is presented in this section. Following the optimized user pairing obtained from the \textbf{($\mathcal{P}$1-A)}, the power allocation is obtained by following the ZOA steps discussed in the previous section. 

\subsection{Subproblem $\mathcal{P}$1-B}
Subsequently, given the fixed value of $\boldsymbol{\Gamma}$, \textbf{U}, the sub-optimization problem for the power allocation matrix (\textbf{$\mathcal{P}$}) can be formulated as:
\begin{equation}
\begin{aligned}
\textbf{($\mathcal{P}$1-B)}:  &\max_{{\textbf{\{P}\}}} \eta_{EE} \\ 
&\text{subject to:} ~ \text{C}_1 ~\text{to} ~\text{C}_3
\end{aligned}
\label{eq12}
\end{equation}
 In power allocation, the candidate solutions are possible power allocation vectors that are updated through ZOA. Finally, the suboptimal solution for the ZOUPPA is obtained, which maximizes the EE for the SGCN. The detailed steps for the proposed ZOUPPA to solve \textbf{($\mathcal{P}$1-B)} are presented in Algorithm  \ref{algo2}.

\subsection{Complexity and convergence Analysis}
\subsubsection{Convergence Analysis}
Let $\mathcal{S} = [\boldsymbol{\Gamma}^*, \boldsymbol{U}^*, \boldsymbol{P}^*]$ denote the optimal solution of \textbf{($\mathcal{P}$1)} after the current iteration, resulting in $\boldsymbol{\eta}_{EE}^*$. The objective function value of \textbf{($\mathcal{P}$1)} at this iteration is represented as $\mathbb{F}(\mathcal{S}_i)$, which corresponds to a subset of the original optimization problem. The objective function value iteratively improves, i.e., $\mathbb{F}(\mathcal{S}_{i+1}) \geq \mathbb{F}(\mathcal{S}_i)$. According to  \cite{mahmood2023joint}, a stable point is reached when the difference between the current and previous iterations is less than $\epsilon$, i.e., $\mathbb{F}_{i+1} - \mathbb{F}_i \leq \epsilon$. In this study, the ZOUPPA converges when the changes in the energy efficiency metric fall below a threshold $\epsilon$, ensuring that the iterative process yields a solution that maximizes energy efficiency in the CR-NOMA-based SGCN.
\subsubsection{Complexity Analysis}
This section provides a worst-case per-iteration complexity analysis for Algorithm \ref{algo2}, designed to iteratively solve the joint optimization problems \textbf{($\mathcal{P}$1-A)} and \textbf{($\mathcal{P}$1-B)}. Specifically, problem ($\mathcal{P}$1-A) involves two decision variables: the channel allocation matrix $\boldsymbol{\Gamma}$ and the user pairing matrix $\boldsymbol{U}$, with $M$ channels and $N$ users, resulting in a per-iteration computational complexity of $\mathcal{O}(MN^2)$. Similarly, problem \textbf{($\mathcal{P}$1-B)} involves the power allocation vector $\textbf{P}$ for $N$ users, leading to a per-iteration complexity of  $\mathcal{O}(N)$. For $i^{max}$ iterations, the overall complexity of Algorithm \ref{algo2} is $\mathcal{O}\left(i^{max}\left(MN^2 + N\right)\right)$, which reflects the iterative process for optimizing both subproblems.
\section{Results and Discussions} \label{RD}
This section assesses the performance of the proposed ZOUP and ZOUPPA for CR-NOMA-based SGCN and presents a comparative analysis of various user pairing and power allocation schemes. The simulations are carried out under the simulation parameters given in Table \ref{tab41}. The simulations are divided into two scenarios: evaluating the performance of ZOUP (single optimization) and ZOUPPA (joint optimization).
\begin{table}[tbph!]
\caption{Simulation Parameters}
\centering
\small
\label{tab41}
\begin{tabular}{cc}
\hline
\textbf{Parameters}           & \textbf{Values} \\ \hline
Number of smart meters (N)            & 100             \\ \hline
Available channels (M)        & 60              \\ \hline
Coverage radius (Rc)          & 100m            \\ \hline
Clusters                     & 0.5 $\times$ SMs         \\ \hline
SNR                           & 30dB            \\ \hline
Path Loss Exponent ($\chi$)   & 2               \\ \hline
Candidate solutions           & 20              \\ \hline
Max iterations                & 100             \\ \hline
DC coordinates                & (0,0)           \\ \hline
Max users in a cluster        & 2               \\ \hline
Channel availability          & random          \\ \hline
$\beta_1$, $\beta_2$       & [0,1]             \\ \hline
Transmission Power of Cluster & 1W              \\ \hline
\end{tabular}
\end{table}
\subsection{Performance evaluation of ZOUP}
In this section, the significance of \textbf{($\mathcal{P}$1-A)} is evaluated and compared with the four benchmark schemes. The OMA considers only single resource allocation to a single user. The other three benchmark schemes are based on NOMA-based pairings: random pairing, adjacent pairing, and user pairing without optimization (UPWO). 
\par 
Random pairing involves pairing users at random without taking into account any particular criteria. In adjacent pairing, the channel gains of the users are sorted, and the users with adjacent channel gains are paired together. Finally, UPWO is the pairing scheme that divides users into two groups according to channel conditions. The users in Group A are the top users with the best channel conditions, and Group B consists of the weak channel users \cite{sarfraz2022capacity}. The UPWO comprises users from both groups, pairing a user exhibiting strong channel characteristics with a weaker user. The performance of ZOUP is analyzed by considering the impact of various parameters, which are given in the following sections.
\subsubsection{Impact of Varying $\beta$ (Power Allocation Constant)}
\begin{figure}[tbph!]
\centering
\includegraphics[width=0.5\textwidth]{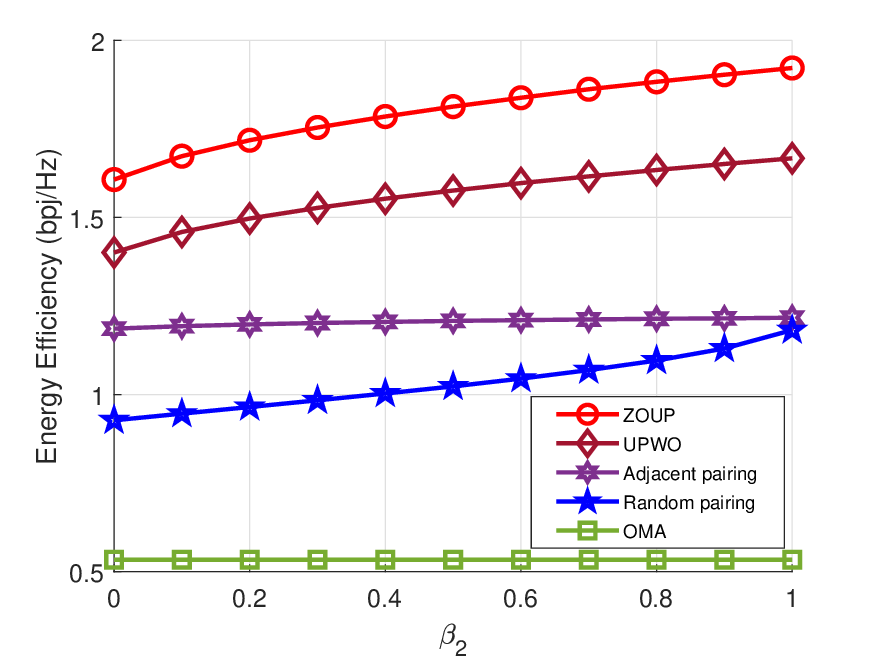}
\caption{Performance comparison between different user pairing schemes with respect to $\beta_2$.}
\label{fig41}
\end{figure}

Figure \ref{fig41} shows the performance comparison of the proposed ZOUP
with the benchmark schemes. The power allocation of \cite{yang2016general}, as indicated in eq. \eqref{eq90a}, clearly demonstrates that modifying the values of the power allocation factors, such as $\beta_1$ and $\beta_2$, significantly impacts the system's performance. The range of values for the constants ($\beta_1$ and $\beta_2$) is between 0 and 1, and their sum is equal to 1. The impact of changing $\beta_2$ from 0 to 1 with an increase of 0.1 is analyzed. The ZOUP demonstrates superiority over the benchmark schemes at both lower and higher values of $\beta_2$. However, increasing the constant $\beta_2$ allocates higher power to strong users, resulting in maximizing the overall EE of the SGCN with higher values of $\beta_2$. The ZOUP achieves an improvement of 13. 27\% in UPWO, 36. 68\% in adjacent pairing, 38. 5\% in random pairing, and 72. 23\% in OMA at $\beta_2$ = 1.


\subsubsection{Impact of Varying SNR}

\begin{figure}[tbph!]
\centering
\includegraphics[width=0.5\textwidth]{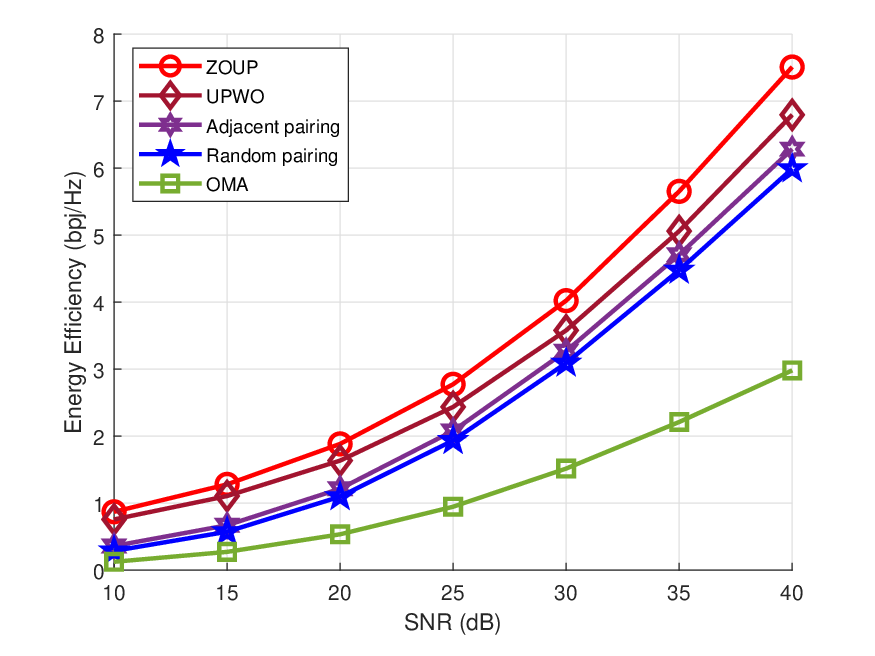}
\caption{Performance comparison between different user pairing schemes at different SNRs}
\label{fig42}
\end{figure}

Higher SNR levels significantly affect the system's performance, leading to greater EE in the SGCN. The higher SNR indicates favorable channel conditions for both strong and weak users, leading to higher overall EE of the SGCN. The performance of ZOUP has been thoroughly evaluated in a range of SNRs, as shown in Figure \ref{fig42}. The SNRs varied from 10 to 40 dB with an increase of 5 dB, and the performance of ZOUP was evaluated. The results clearly show that ZOUP performs better at lower SNR levels (10 dB) and excels with other pairing schemes at higher SNRs (40 dB). The ZOUP outperforms UPWO and produces an improvement of 13.11\%. It achieves 58.67\% for adjacent pairing, 67.14\% for random pairing, and 85.84\% for OMA at 10 dB of SNR. Similarly, the ZOUP provides a 9. 5\% improvement in UPWO, 16. 32\% to adjacent pairing, 20. 21\% to random pairing and 60. 31\% to OMA at 40 dB of SNR. This indicates that the ZOUP provides a better enhancement to the benchmark pairing schemes even at lower SNRs and achieves higher energy efficiency at higher SNRs.

\subsubsection{Impact of Diverse Environments}
\begin{figure}[t]
\centering
\includegraphics[width=0.5\textwidth]{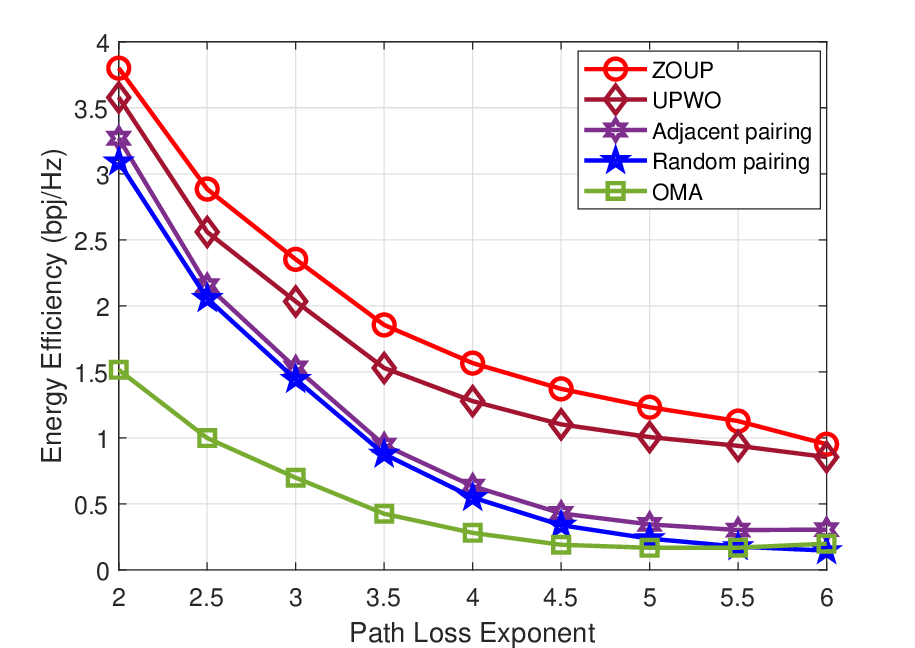}
\caption{Impact of various environments.}
\label{fig43}
\end{figure}

In denser environments, the system's performance is significantly degraded due to increased multipath fading, shadowing, and other losses. These adverse conditions lead to increased signal attenuation and degradation, which negatively impact EE. To investigate the impact of different environmental conditions on system performance, we examine the path loss exponent $\chi$ for various environments such as rural, suburban, urban, and dense urban areas. Figure \ref{fig43} illustrates how the path loss exponent affects the ZOUP and benchmark pairing schemes in different environments.
The simulations represent the performance improvement of 17.57\% to UPWO, 49\% to adjacent pairing, 52.72\% to random pairing, and 77.08\% to OMA at $\chi$=3.5. The ZOUP demonstrated superiority over UPWO, indicating a 16.64\% improvement. It achieved 73.24\% with adjacent pairing, 84.38\% with random pairing, and 85.1\% with OMA at a $\chi$= of 5.5.

\subsubsection{Impact of Varying User Density}
\begin{figure}[tbph!]
\centering
\includegraphics[width=0.5\textwidth]{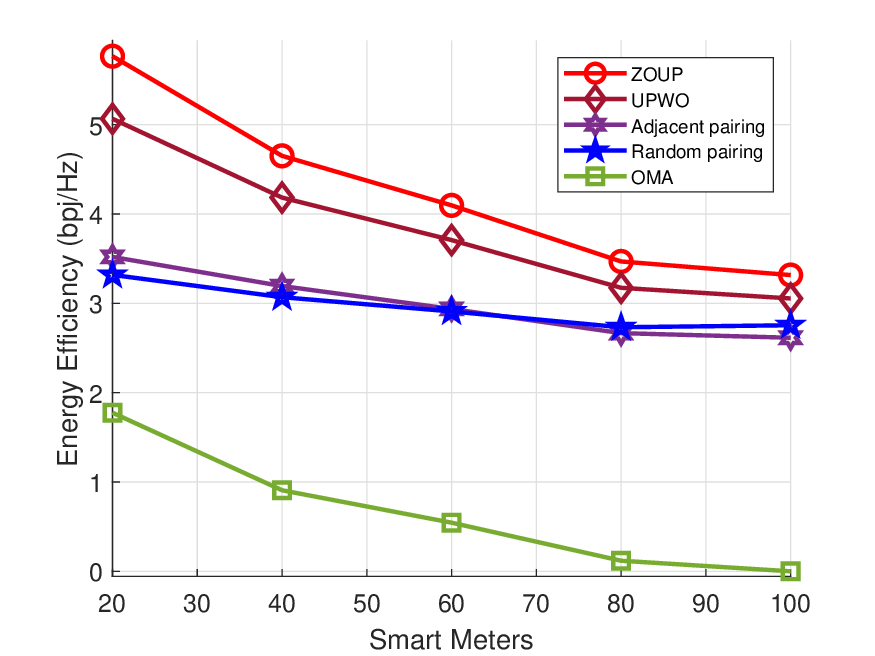}
\caption{Performance comparison for increasing SMs.}
\label{fig44}
\end{figure}

Lower user (SM) density while fixing channel availability increases the probability of selecting the channel with good conditions. Increased user density affects the SGCN's EE when channel availability is fixed because there will be fewer options for selecting favorable channels. 
The EE decreases as user density increases, as shown in Figure \ref{fig44}. The performance improvement stands at 9.5\% for UPWO, 28.31\% for adjacent pairing, 29\% for random pairing, and 86.71\% for OMA with 60 users. The adjacent pairing starts losing performance compared to the random pairing, and the performance of OMA starts vanishing when SMs exceed 60. Similarly, the ZOUP achieves an 8\% improvement, adjacent pairing delivers a notable 16.31\% improvement, random pairing yields a substantial 21.14\% improvement, and OMA produces an outstanding 87.94\% improvement with 100 users. This clearly illustrates ZOUP's superior performance compared to other benchmarks across a wide range of user densities.

\subsubsection{Impact of Channel Availability}
\begin{figure}[b]
\centering
\includegraphics[width=0.5\textwidth]{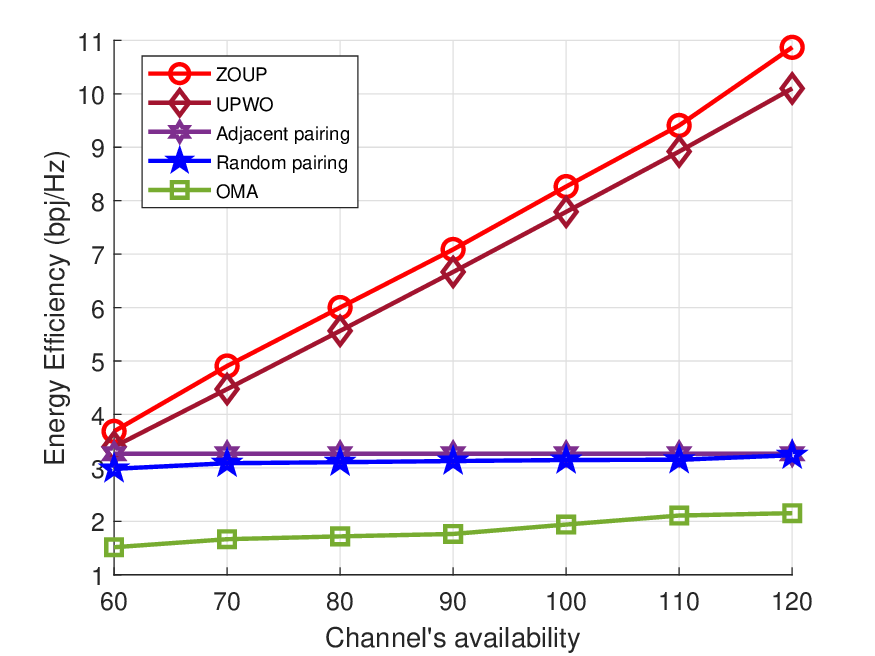}
\caption{Performance comparison for increasing the channel availability.}
\label{fig45}
\end{figure}

Channel availability plays a vital role in SGCN's EE performance when the number of channels increases. Higher channel availability means more channels with better characteristics are available, and pairing these channels results in a higher EE. In contrast, limited channel availability means there are fewer options available for pairing, which reduces the potential for increasing the EE.
Figure \ref{fig45} illustrates the performance comparison of ZOUP with benchmarks as the channel availability increases from 60 to 120 channels.
The simulation results indicate that ZOUP improves UPWO by 7.82\%, adjacent pairing by 11.37\%, random pairing by 19.06\%, and OMA by 58.9\% across 60 channels. Likewise, the ZOUP achieves a 5.72\% improvement over UPWO, 60.5\% over adjacent pairing, 61.92\% over random pairing, and 76.53\% over OMA with 100 channels.  This proves that the ZOUP achieves higher performance when the channel availability is either low or high.

\subsubsection{Impact of increasing coverage radius ($R_{C}$)}
\begin{figure}[t]
\centering
\includegraphics[width=0.5\textwidth]{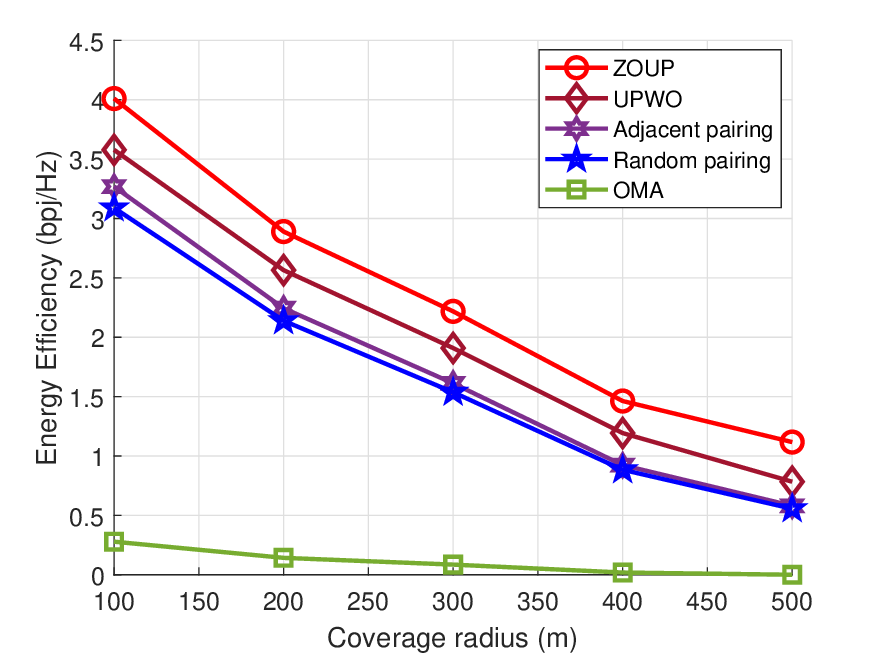}
\caption{Impact of an increasing coverage region.}
\label{fig46}
\end{figure}

As the signal travels from the transmitter to the receiver, it loses some of its power due to factors such as multipath fading, shadowing, and path loss, which are essential among those factors. The path loss increases as the distance between the transmitter and the receiver increases. This degrades the quality of the received signal and makes it difficult for the decoder to recover the signal. Increasing the coverage radius extends the transmission distance between the SMs and the NAN-DC. Therefore, the overall EE of the SGCN will degrade as the coverage region increases. In Figure \ref{fig46}, the ZOUP shows a performance improvement of 10. 77\% compared to UPWO, 18.48\% compared to adjacent pairing, 22.94\% compared to random pairing and 93. 05\% compared to OMA at a coverage radius of 100 m. Similarly, the ZOUP achieves a 29.86\% improvement in UPWO, a 48.03\% improvement in adjacent pairing, a 50.22\% improvement in random pairing and a 98\% improvement in OMA at the coverage radius of 500m. This highlights the effectiveness of ZOUP compared to other pairing schemes, both at lower and larger coverage radii.

\subsection{Joint User Pairing and Power Allocation (ZOUPPA)}
The suboptimal user pairs obtained from \textbf{$\mathcal{P}$1-A} are allocated power through ZOA, which significantly increases the EE of the SGCN.
The performance of the joint user pairing and power allocation (ZOUPPA) for \textbf{($\mathcal{P}$1-B)} is evaluated in this section, where ZOUPPA (joint optimization) is compared with ZOUP (single optimization) and UPWO (without optimization). Table \ref{tab3} shows the significant improvement gained from optimizing user pairing and power allocation instead of solely optimizing user pairing.
The performance of the ZOUPPA is evaluated for five different parameters, i.e., SNR, path loss exponent ($\chi$), user density ($SM_{S}$), channel availability ($M$), and coverage radius ($R_{C}$).
\par
Table \ref{tab3} indicates that ZOUPPA achieves significantly higher performance, with a 25.39\% improvement over ZOUP and a 53.25\% improvement over UPWO at 15 dB SNR. In addition, it shows a 13.32\% enhancement for ZOUP and a significant 34.35\% improvement for UPWO at 30 dB SNR. At $\chi$ = 3, ZOUPPA shows an 18.77\% improvement over ZOUP and a 44.59\% improvement over UPWO. Furthermore, at $\chi$  = 4, it demonstrates a 28.32\% improvement against ZOUP and an outstanding 57.86\% improvement against UPWO. The higher user density has a greater impact on EE performance. At 200 SMs, ZOUPPA achieves 52.4\% higher EE than ZOUP and 57.69\% higher EE than UPWO. Increasing the user density to 300 SMs results in 17.29\% more than ZOUP and 24.39\% more than UPWO, showcasing the greater scalability of the ZOUPPA. The availability of more channels significantly increases the likelihood of discovering channels with fair conditions. As a result, the EE for the ZOUPPA is 18.77\% higher than that for ZOUP and 53.32\% higher than that for UPWO at 50 channels. Similarly, with 100 channels, the EE improves to 27.06\% for ZOUP and 61.2\% for UPWO. When the coverage radius of NAN-DC is expanded, performance decreases noticeably, leading to a decline in EE. Expanding the coverage radius of NAN-DC has been observed, which leads to a noticeable decrease in performance, resulting in a decline in EE. The performance of ZOUPPA has been evaluated across various coverage areas, demonstrating a 32.6\% increase in EE compared to ZOUP and a significant 61.48\% improvement in UPWO compared to UPWO at 200m. When extending the coverage to 400m, there was a 35.66\% improvement compared to ZOUP and 71.46\% to UPWO.
\begin{figure}[t]
\centering
\includegraphics[width=0.5\textwidth]{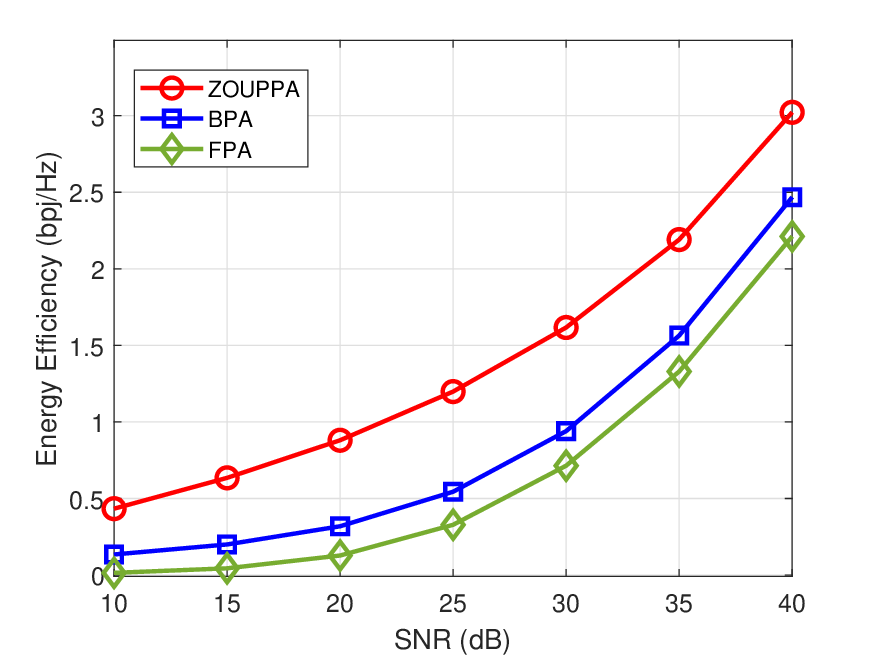}
\caption{Impact of an increasing SNR for a joint optimization problem.}
\label{fig48}
\end{figure}
\begin{table*}[tbph!]
\setlength{\tabcolsep}{10pt} 
\caption{Performance Comparison of ZOUPPA with ZOUP and UPWO}
\label{tab3}
\small
\centering
\begin{tabular}{ccccccc}
\hline
\multicolumn{2}{c}{\textbf{Parameters}} & \textbf{UPWO} & \textbf{ZOUP (P1-A)} & \textbf{ZOUPPA (P1-B)} & \textbf{\% Improvement to UPWO} & \textbf{\% Improvement to ZOUP} \\ 
\hline
\multirow{2}{*}{$SNR$} & 15 dB & 0.4783 & 0.7633 & 1.0231 & 53.25\% & 25.39\% \\
                       & 30 dB & 1.7382 & 2.2950 & 2.6476 & 34.35\% & 13.32\% \\
\hline
\multirow{2}{*}{$\chi$} & 3    & 1.1029 & 1.6168 & 1.9905 & 44.59\% & 18.77\% \\
                        & 4    & 0.6672 & 1.1349 & 1.5832 & 57.86\% & 28.32\% \\
\hline
\multirow{2}{*}{$SMs$} & 200   & 0.9709 & 1.0925 & 2.2950 & 57.69\% & 52.40\% \\
                       & 300   & 0.8086 & 0.8854 & 1.0694 & 24.39\% & 17.21\% \\
\hline
\multirow{2}{*}{$M$}   & 50    & 1.0052 & 1.7491 & 2.1534 & 53.32\% & 18.77\% \\
                       & 100   & 1.7079 & 3.2109 & 4.4021 & 61.20\% & 27.06\% \\
\hline
\multirow{2}{*}{$R_{C}$} & 200m  & 0.3947 & 0.6906 & 1.0246 & 61.48\% & 32.60\% \\
                         & 400m  & 0.0830 & 0.1871 & 0.2908 & 71.46\% & 35.66\% \\
\hline
\end{tabular}
\end{table*}

\par
The performance of the proposed scheme is assessed by allocating power to suboptimal user pairs using three distinct power allocation schemes. The first scheme, fixed power allocation (FPA), employs preset power allocation coefficients $\delta_{v}$=0.75 and $\delta_{u}$=0.25 for user 1 and user 2, respectively. The second scheme, constant beta-based power allocation (BPA) from \cite{yang2016general}, calculates the power allocation coefficients using eq. \eqref{eq90a} for user 1 and user 2. The third scheme, ZOUPPA, is the proposed power allocation method that calculates power allocation coefficients using ZOA. 
Figure \ref{fig48} shows that ZOUPPA yields an 18.34\% improvement over BPA and a 26.82\% enhancement over FPA at 40dB SNR.

\section{Conclusion} \label{Con}
The paper presents a novel approach to optimizing the EE of a CR-NOMA-based SGCN by addressing a joint user pairing and power allocation problem. By incorporating CR to tackle spectrum scarcity and integrating NOMA, the network's EE is significantly improved. The formulated problem for user pairing is first solved using ZOA, and comparisons with benchmarks illustrate the effectiveness of the ZOUP. The suboptimal user pairs extracted from the ZOUP are then assigned power through ZOA, which is named ZOUPPA. The performance of the ZOUPPA was assessed using a range of parameters, including SNR, x, SMs, M, and Rc. The results demonstrate that ZOUPPA outperforms ZOUP and UPWO across different combinations of these parameters. This highlights the superior performance of ZOUPPA in scenarios with lower SNRs, suburban to densely urban areas, enhanced scalability, fluctuating channel availability, and diverse coverage regions. Additionally, the effectiveness of ZOUPPA across various power allocation methods is verified, highlighting its superior performance in improving the SGCN's EE. In conclusion, ZOUPPA is identified as the more effective joint optimization approach than the benchmarks. ZOUPPA achieved an 18.34\% improvement over the power allocation scheme of \cite{yang2016general} and a 26.82\% improvement over fixed power allocation at 40 dB SNR.

	\bibliographystyle{IEEEtran}
	\bibliography{ReferenceBibFile}
	
\end{document}